\documentclass[sigconf]{acmart}
\usepackage{algorithm}
\usepackage{algpseudocode}
\usepackage{makecell} 
\usepackage{enumitem}
\usepackage{float}
\usepackage{booktabs}
\usepackage{graphicx}
\usepackage{arydshln}
\usepackage{tablefootnote}
\usepackage[flushleft]{threeparttable}

\AtBeginDocument{%
  }

\setcopyright{acmlicensed}
\copyrightyear{2025}
\acmYear{2025}
\acmDOI{XXXXXXX.XXXXXXX}

\acmConference[Conference acronym 'XX]{Make sure to enter the correct
  conference title from your rights confirmation emai}{June 03--05,
  2018}{Woodstock, NY}

\acmISBN{978-1-4503-XXXX-X/18/06}

\begin{document}

\author{Jungbae Park}
\affiliation{
  \institution{Bucketplace Inc.}
  \city{Seoul}
  \country{Republic of Korea}
}
\email{jungbae.park@bucketplace.net}

\author{Heonseok Jang}
\affiliation{
  \institution{Bucketplace Inc.}
  \city{Seoul}
  \country{Republic of Korea}
}
\email{hs.jang@bucketplace.net}

\title{MOHPER: Multi-objective Hyperparameter Optimization Framework for E-commerce Retrieval System}

\begin{abstract}

E-commerce search optimization has evolved to include a wider range of metrics that reflect user engagement and business objectives. Modern search frameworks now incorporate advanced quality features, such as sales counts and document-query relevance, to better align search results with these goals. Traditional methods typically focus on click-through rate (CTR) as a measure of engagement or relevance, but this can miss true purchase intent, creating a gap between user interest and actual conversions. Joint training with the click-through conversion rate (CTCVR) has become essential for understanding buying behavior, although its sparsity poses challenges for reliable optimization. This study presents MOHPER, a Multi-Objective Hyperparameter Optimization framework for E-commerce Retrieval systems. Utilizing Bayesian optimization and sampling, it jointly optimizes both CTR, CTCVR, and relevant objectives, focusing on engagement and conversion of the users. In addition, to improve the selection of the best configuration from multi-objective optimization, we suggest advanced methods for hyperparameter selection, including a meta-configuration voting strategy and a cumulative training approach that leverages prior optimal configurations, to improve speed of training and efficiency. Currently deployed in a live setting, our proposed framework substantiates its practical efficacy in achieving a balanced optimization that aligns with both user satisfaction and revenue goals.
\end{abstract}



\maketitle

\section{Introduction}


With the growth of online shopping in modern society, e-commerce platforms now account for a vast number of user queries daily. The product search engines within these platforms play a pivotal role, acting as arbitrators that match consumer intent with product listings. Beyond simply interpreting and retrieving relevant items, e-commerce search engines serve as the primary conduit between digital vendors and consumers, shaping the user experience and influencing engaging and purchasing behavior across online markets. To improve both the engagement of consumers and the profitability of vendors, e-commerce search optimization has rapidly progressed from simple keyword matching to sophisticated, multi-feature frameworks that integrate relevance, user engagement, and business-oriented metrics \cite{wu2018LETORIFClickIntoPurchase, zheng2022MOPersonalizedRetrievalTaobaoSearch, karmaker2017LTRforEcommerceSearchLETEOR}.

Traditional search systems often rely on lexical retrieval techniques such as TF-IDF and BM25 \cite{ramos2003tf-idf, robertson2009bm25}, which match keywords between user queries and product documents. More recently, neural retrieval models based on large language models (LLMs) \cite{arevalillo2013MOInImageRetrieval, li2023adaptiveHyperparamterLearningForDeepSemanticRetrieval, bhagdev2008hybridsearch} and multimodal embeddings have improved semantic alignment \cite{radford2021clip}, enabling richer query-document representations. Although most previous studies focus on relevance scores of the hybrid retrieval using both BM25 and dense retrieval, these similarity scores do not imply the preferences, interests, or engagements of users, which are crucial for e-commerce, generally available by interpreting user interaction logs.
Click-based optimization, especially CTR, has become a standard proxy for user preference. Yet, CTR is an imperfect signal: it overestimates interest and underrepresents conversions. This mismatch motivates the use of more direct metrics such as click-through-conversion rate (CTCVR), which captures actual purchase intent. However, conversion is inherently sparse and biased, since it only exists for a subset of clicked items, risking overfitting to a small subset of items \cite{zhang2022ctnocvr, ma2018EntireSpaceMTforEstimatingPostClickConversion}. Traditional retrieval or ranking methods, including BM25 \cite{robertson2009bm25}, dense vector retrievals \cite{wang2021bertwithBM25, li2023adaptiveHyperparamterLearningForDeepSemanticRetrieval, li2023AdaptiveHPLearningforSemanticRetrieval}, and learning-to-rank (LTR) 
 \cite{burges2005ltr} typically optimizes for single objectives, which can result in misalignment between relevance, user engagement, and conversion. This single-objective focus limits adaptation to rapidly changing e-commerce trends and degrades performance in online environments \cite{karmaker2017LTRforEcommerceSearchLETEOR}.

\begin{figure*}[!thb]
  \centering
  \includegraphics[width=.85\linewidth]{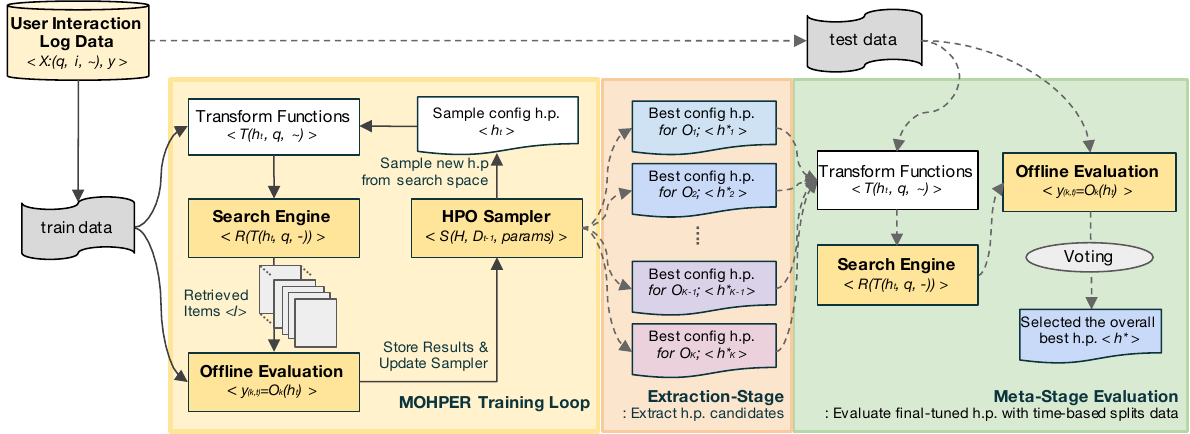}
  \caption{Overall architecture diagram of the proposed framework, MOHPER.}
  \Description{architecture}
  \label{Architecture}
\end{figure*}

In response, multi-objective strategies are gaining traction in e-commerce, aiming to optimize both CTR and CTCVR to align engagement with purchase intent. This study presents MOHPER, a multi-objective hyperparameter optimization (MOHPO) framework designed for e-commerce search. Using Bayesian optimization as \cite{li2018BOForRetrieval} and advanced sampling techniques, MOHPER jointly optimizes CTR and CTCVR, offering a balanced solution to enhance both engagement and conversions in selecting optimal hyperparameters (HPs). We systematically evaluate sampling strategies within MOHPER to ensure robust multi-metric optimization, addressing the demand for adaptive, multi-objective solutions in e-commerce.
The main contributions of this paper include:

\begin{enumerate}[label=\roman*.]
    \item introduction of a novel hyperparameter optimization (HPO) framework specifically tailored for e-commerce search services, focusing on objectives such as CTR and CTCVR, essential indicators of user engagement and purchase intent. This is the first application of a multi-objective HPO framework in e-commerce retrieval and ranking tasks.
    \item suggestion for advanced methodologies for HP selection, including a meta-configuration voting strategy and a cumulative training that leverages prior optimal configurations.
    \item thorough testing of the MOHPER framework using various samplers and search scenarios, validated through extensive A/B testing with a significant dataset from OHouse, one of South Korea's leading furniture e-commerce platforms.
\end{enumerate}

\section{MOHPER: Problem Formulation and Design}

In this section, we present the proposed MOHPER framework. We begin by formulating the multi-objective problem within retrieval systems and provide an overview of the HPO algorithm. Subsequently, we identify the relevant objectives for joint optimization in the e-commerce setting. To enhance performance during the meta-evaluation stage and to select appropriate configurations in the multi-objective optimization context, we introduce two methods: meta-configuration voting and cumulative HP stacking. The proposed methods are illustrated in Figure \ref{Architecture}.

\subsection{Multi-Objective HPO in Retrieval Systems}

\begin{figure}[!thb]
  \centering
  \includegraphics[width=.77\linewidth]{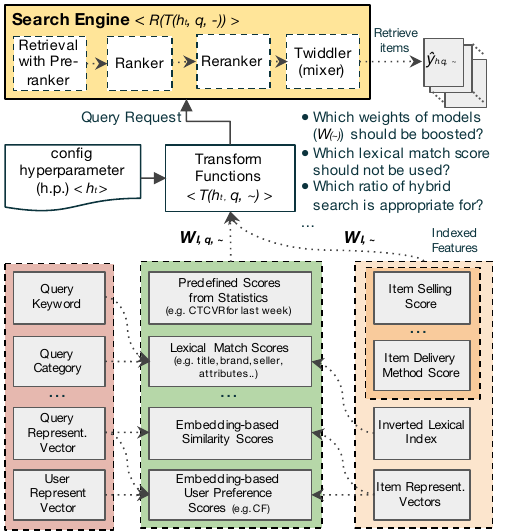}
  \caption{Transform function to modulate query request for a search engine as the change of config HP}
  \label{ModelInput}
  \Description{ModelInput}
\end{figure}

Formally, the goal is to identify a set of HPs ($h \in \mathcal{H}$) that maximizes the performance of a retrieval engine across multiple objectives $O = \{O_1, O_2, \dots, O_M\}$ simultaneously. Given an observation dataset $\mathcal{D}_0 = \emptyset$ and a set of ground truth labels $Y = \{y_{1, (\cdot)}, y_{2, (\cdot)}, \dots, y_{m, (\cdot)}\}$ for each objective, the optimization seeks to sample and evaluate HP configurations to approximate optimal performance iteratively. In other words, the search system is evaluated by selecting an HP configuration $h_t$ from the search space $\mathcal{H}$ at each trial $t = 1, 2, \dots, N$. During optimization progress, various sampler methods, such as random search, grid search, or Bayesian optimization (e.g., tree-structured Parzen estimator (TPE) or Gaussian process (GP), etc.), can be employed to sample HP configurations.
At each trial, $t$, an HP configuration $h_t$ is sampled, and the retrieval system is evaluated on a set of queries. The configuration $h_t$ influences the retrieval process through transform functions that adapt the retrieval engine's behavior based on the sampled HPs. This could involve altering ranking strategies, modifying relevance weights, or adjusting search heuristics (see Figure \ref{ModelInput}).
Specifically, the transform function can be expressed as $ T(h_t, q) $, which represents the transformed retrieval process based on the HP configuration $h_t$ applied to query $q$. Depending on the sampled HPs, this transformation could modify how items are ranked or filtered for search engines.
For each query $q$, the system produces predictions (ranked items), and the relevance/quality metrics for these items are evaluated using objectives, respectively. The metric objective $m$ can be computed based on the predicted rankings $\hat{y}_{h_t, q, (\cdot)}$ and compared to the ground truth labels $y_{m, q, (\cdot)}$, formulated as $z_{t,m,q} = O_k(y_{m, q, (\cdot)}, \hat{y}_{h_t, q, (\cdot)})$. Performance evaluations across all queries are then aggregated to produce a performance metric for each objective. The aggregated metrics are used to update the optimization process, guiding the search for better HP configurations. This process continues iteratively until convergence, where the goal is to select a HP configuration that optimizes the trade-off between the multiple objectives. The overall algorithm with TPE sampler is summarized in Algorithm \ref{multi_objective_general}.

\begin{algorithm}[htp!]
\small
\center
\caption{MOHPER Training with the TPE Sampler Example}\label{multi_objective_general}
\begin{algorithmic}[1]
\State Initialize HPO study with sampler $S$ (e.g., TPE, GP, \dots)
\State Define objectives $O = \{O_1, O_2, \dots, O_M\}$, with corresponding ground truth labels $Y = \{y_{1, (\cdot)}, y_{2, (\cdot)}, \dots, y_{M, (\cdot)}\}$
\State If it is in cumulative training, load previous top observations into datasets  $\mathcal{D}_0 = \{h_{prev\ A}, z_{prev\ A, 1}, z_{prev\ A, 2}, \cdots \}$ in cumulative training, else initialize observation dataset $\mathcal{D}_0 = \emptyset$

\For{each trial $t = 1, 2, \dots, N$}
    \State Sample hyperparameters (HPs):
    $$ h_t \sim S(\mathcal{H}, \mathcal{D}_{t-1}, \text{params}) $$
    
    \For{each query $q = q_1, q_2, \dots, q_L$}
        \State Retrieve Items from Search Engine ($R$) with Transform Function ($T$):
        $$ \hat{y}_{h_t, q, (\cdot)} = R(T(h_t, q, (\cdot))) $$
        
        \State Evaluate objectives:
        $$ z_{t,m,q} = O_m(y_{m, q, (\cdot)}, \hat{y}_{h_t, q, (\cdot)}), \quad \forall O_m \in O $$
    \EndFor
    \State Aggregate evaluated objectives (e.g., average):
    $$ z_{t,m} = Agg(z_{t,m,q}), \quad \forall O_m \in O $$
    
    \State Store results $(h_t, z_{t,1}, z_{t,2}, \dots, z_{t,M})$

    \State \textbf{Update TPE sampler state based on evaluations:}
    \State Define threshold $\gamma_{t, m}$ on each objective, for $v$ quantile:
    $$ \gamma_{t, m} = \text{quantile}(z_{t,m}, v), \quad \forall O_m \in O $$

    \State Update good and bad HP distributions using $\gamma_{t, m}$:
    $$ l_m(\mathbf{h}) = p(\mathbf{h} \mid z_{t,m} \leq \gamma_{t, m}), \quad g_m(\mathbf{h}) = p(\mathbf{h} \mid z_{t,m} > \gamma_{t, m}) $$
    
    \State Calculate expected improvement (EI) for candidate HPs and select HPs that maximize EI:
    $$ \text{EI}(\mathbf{h}) = \prod_{m=1}^M \frac{l_m(\mathbf{h})}{g_m(\mathbf{h})}, \quad h_t = \arg\max_{\mathbf{h} \in \mathcal{H}} \text{EI}(\mathbf{h}) $$
    
    \State \textbf{Update dataset if improvement is achieved:} 
    If $\text{EI}(h_t) > 0$, then update the observation dataset:
    $$ \mathcal{D}_t = \mathcal{D}_{t-1} \cup \{(h_t, z_{t,1}, z_{t,2}, \dots, z_{t,M})\} $$

\EndFor
\end{algorithmic}

\end{algorithm}

\subsection{Joint Optimization in E-Commerce Context} \label{joint_optimization_in_e_commerce}

\begin{figure}[!tb]
  \centering
  \includegraphics[width=.66\linewidth]{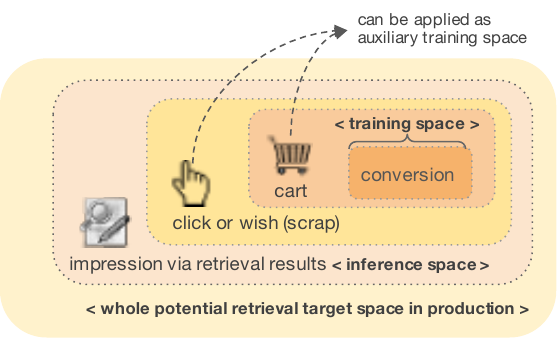}
  \caption{Illustration of the e-commerce conversion funnel starting from item impressions, highlighting issues of sample selection bias and data sparsity in user interaction logs}
  \Description{Illustration of sample selection bias and data sparsity issue}
  \label{label_sparsity}
\end{figure}

To enhance user engagement and sales in e-commerce while addressing sample selection bias and data sparsity, it is essential to mine user interaction logs, including clicks and conversions per impression. Although MOHPO is designed to optimize any metrics of objectives as other black-box HPO algorithms do, in this study, we track the retrieval performance through the following objectives:
\begin{enumerate}
    \item Click-Through Conversion Rate (CTCVR): $\#\text{purchases}\ / \#\text{imp.}$
    \item Click-Through Rate (CTR): $\#\text{clicks}\ / \ \#\text{imp.}$
    \item Click-Through Cart-Adding Rate (CTCAR): $\#\text{carts}\ / \ \#\text{imp.}$
\end{enumerate}
Here, $\#\text{imp.}$ represents the number of item impressions in retrievals. CTCVR is the primary objective, with CTR, and CTCAR serving as auxiliary objectives to mitigate selection bias and sparsity (Figure \ref{label_sparsity}) \cite{ma2018EntireSpaceMTforEstimatingPostClickConversion, karmaker2017LTRforEcommerceSearchLETEOR}.
Specifically, For any query $q$ and trial $t$, performance metrics $z_{t, q, (\cdot)}$ (e.g. precision@K, nDCG@K, mAP@K, and recall@K) are computed when ground truth labels  $y_{q, (\cdot)}$ are available and the top-K items are retrieved ($ \hat{y}_{h_t, q, (\cdot)}$). Labels for these items are derived from objective-specific statistics (e.g. $\# imp_{q, i}$, $\# click_{q, i}$, $\# purchase_{q, i}$), with thresholding for positive/negative scoring at the corresponding objective. Also, while naive statistical rates are available, filtering based on impression counts and Bayesian smoothing for rare events can further refine these scores \cite{wang2011bayesiansmoothedCTR}. 
Following metric calculation $z_{t,m,q}$, metrics are aggregated across queries, and MOHPO proceeds with multi-objective tuning to enhance engagement, conversion efficiency, and purchase intent, advancing retrieval performance in e-commerce.

\subsection{Hyperparameter Extraction Stage}

Upon completing HPO training, the framework selects a diverse set of candidate HP configurations based on their performance across multiple objectives, such as training targets or a weighted metric sum (see the middle stage of Figure \ref{Architecture}). These top configurations are then stored for further meta-level evaluation to estimate online performance and for use in subsequent cumulative training.

\subsection{Meta Stage Evaluation and HP selection }

To mitigate overfitting from training data, we introduce a meta-evaluation stage that validates top HP configurations on unseen queries and time-based splits, improving robustness and generalization~\cite{cawley2010over,shani2011evaluating}. Each configuration is evaluated across multiple objectives, including training and other service-relevant objectives.

We introduce meta-configuration voting, based on a voting-based selection strategy~\cite{chan1995voting}, where each $h \in \mathcal{H}_{\text{top}}$ is scored as:
\begin{equation}
\text{Vote}(h) = \sum_{k=1}^{K} \mathbb{1}\{h\in\text{top-}n \text{ on objective } k \}, \quad
h^* = \arg\max_{h} \text{Vote}(h).
\end{equation}
This ensemble-style approach reduces variance~\cite{breiman1996bagging} and aligns with stability-based generalization theory~\cite{bousquet2002stability}. Consistent performance across independent validation splits further bounds generalization error via concentration inequalities such as the Chernoff bound~\cite{chernoff1952measure}.

\subsection{Cumulative Training}
\label{cumulative_training}

To enhance convergence and sample efficiency, we adopt a cumulative training strategy by reusing top configurations from previous rounds:
\begin{equation}
\mathcal{D}_0 = \{(h, z_1, \dots, z_M) \mid z_m \geq \gamma_m \ \forall m \in \mathcal{O} \}.
\end{equation}
This warm-started optimization avoids redundant exploration and accelerates convergence~\cite{lin1992self}.
Cumulative training can be interpreted through an empirical Bayes lens~\cite{morris1983parametric}, where prior high-performing configurations act as data-driven priors. This improves algorithmic stability and reduces sensitivity to distribution shifts~\cite{bousquet2002stability}. Given the search history $S_{t-1}$, the candidate pool at step $t$ becomes:
\begin{equation}
S_t = S_{t-1} \cup \{h_t\},
\end{equation}
allowing knowledge accumulation over iterations and more efficient exploration of the search space.

\section{Experiments}

We conduct extensive experiments to evaluate the effectiveness and the performance of the MOHPO framework and its components, and the following research questions (RQs) are answered:

\begin{enumerate}[label=\textbf{RQ \arabic*}.]
\item Multi-Objective Balancing Across Objectives and Samplers: How effectively does MOHPO balance multi-objective metrics (e.g., CTR, CTCVR, CTCAR) under varying objective configurations and HPO sampling strategies? 

\item Hybrid Search Optimization with MOHPO: How well does MOHPO integrate hybrid search mechanisms to optimize for both relevance and multi-objective goals, enhancing user engagement and conversion?

\item Effectiveness of Meta-Configuration Voting and Cumulative Training: Do meta-configuration voting and cumulative training improve MOHPO’s generalizability on unseen queries and adaptability to temporal shifts? Are these strategies critical for maintaining high performance in dynamic e-commerce environments?

\item Online Deployment and Practical Feasibility:
Can MOHPO be effectively deployed in real-world online services while ensuring computational efficiency and demonstrating tangible improvements in retrieval/ranking performance?

\end{enumerate}
\subsection{Datasets and Experimental Setups}

\subsubsection{Datasets and Split Settings} \label{datasets_and_experimental_setups}

For offline evaluation, we utilize user interaction logs from the OHouse\footnote{https://ohou.se/} platform, capturing diverse engagement events such as clicks, purchases, and cart additions across a broad spectrum of queries. In this study, we draw a subset of interaction logs from users in the Republic of Korea spanning from June 16 to September 15, 2024, for training data. The subsequent month’s logs (September 16 to October 15, 2024) are employed for meta-stage evaluation. Offline evaluation is conducted across two scenarios: (1) a keyword-search scenario, where queries are based on textual input, and (2) a retrieval scenario, where queries are formulated through product category IDs, representing groups such as furniture, electronics, and kitchen goods. The statistics of the training and evaluation datasets are summarized in Table \ref{statistic_of_datasets}.

\begin{table}[H]
\centering
\caption{Dataset statistics used in training and evaluation.}
\resizebox{.85\columnwidth}{!}{%
\label{statistic_of_datasets}
\small
\begin{tabular}{l|cccc}
\toprule
\textbf{Split} & \# Queries & \# Impressions & \# Clicks & \# Conversions \\
\midrule
SERP Train     & 1,000      & 367.54M     & 12.02M      & 1.21M          \\
SERP Test      & 5,000      & 377.37M     & 11.98M      & 1.23M          \\
CPLP Train     & 1,379      & 436.76M     & 9.58M      & 0.86M          \\
CPLP Test      & 1,255      & 106.48M     & 2.59M      & 0.23M          \\
\bottomrule
\end{tabular}
}
\end{table}

\subsubsection{Retrieval Engine and Indexing Setting}

We use Elasticsearch v8.14 for offline experimentation and v8.8 in production. Search indices consist of both sparse and dense embeddings, supporting lexical and semantic retrieval. Dense indexing uses HNSW \cite{malkov2018hnsw_indexing}, and sparse retrieval uses standard inverted indices \cite{salton1983inverted_indexing}.

Each query $q$ is issued to the search engine $R$ via a transform function $T(h, q)$ that modifies the engine's behavior according to the HP configuration $h$. For efficiency, evaluations are batched using Elasticsearch’s multi-search API\footnote{https://www.elastic.co/guide/en/elasticsearch/reference/current/search-multi-search.html} to execute concurrent searches within a single request. This concurrent processing is essential for the HPO process, where search results are repeatedly requested to generate predictions based on various HP configurations, significantly improving training efficiency through parallelization.

\subsubsection{Hyperparamter Optimization Framework}

While various HPO frameworks can be applied to MOHPER, we utilize Optuna  \cite{akiba2019optuna}, enabling the evaluation of diverse samplers and configuration settings. Optuna’s flexibility and adaptability make it well-suited for the rigorous testing of multiple HPO strategies within MOHPER.
To address the need for concurrent and parallel processing across multiple training trials, we integrate RayTune\footnote{https://docs.ray.io/en/latest/tune/index.html}, built on the Ray framework\footnote{https://github.com/ray-project/ray}, which is designed to support scalable, distributed computing \cite{liaw2018raytune}. This integration supports high-throughput, parallelized experimentation, enhancing the efficiency of testing various sampler configurations within MOHPER.

\begin{table*}[h!]
\centering
\caption{Performance comparison of various models and optimization objectives on SERP test data. For sampling-based HPO (excluding grid search), 100 samples are limitedly used by default to show sample efficiencies unless otherwise specified.}
\label{SERP_offline_result}
\centering
\resizebox{.85\textwidth}{!}{%
\centering
\begin{tabular}{l||cccc:cccc|c}
\toprule
\textbf{Method} & \makecell{\textbf{CTR} \\ \textbf{nDCG@20}} & \makecell{\textbf{CTR} \\ \textbf{nDCG@100}} & \makecell{\textbf{CTR} \\ \textbf{prec@100}} & \makecell{\textbf{CTR} \\ \textbf{avg metrics}} & \makecell{\textbf{CTCVR} \\ \textbf{nDCG@20}} & \makecell{\textbf{CTCVR} \\ \textbf{nDCG@100}} & \makecell{\textbf{CTCVR} \\ \textbf{prec@100}} & \makecell{\textbf{CTCVR} \\ \textbf{avg metrics}} & \makecell{\textbf{Overall} \\ \textbf{avg metrics}} \\
\midrule
BM25 & 0.2273 & 0.2630 & 0.2724 & 0.2542 & 0.0779 & 0.1470 & 0.3977 & 0.2076 & 0.2309 \\
ANN (KoBERT) & 0.5295 & 0.3592 & 0.2836 & 0.3908 & 0.2799 & 0.3032 & 0.4027 & 0.3286 & 0.3597 \\
ANN (Jina-clip-v2, query to item text) & 0.4071 & 0.3206 & 0.2792 & 0.3356 & 0.2217 & 0.2643 & 0.4349 & 0.3070 & 0.3213 \\
ANN (Jina-clip-v2, query to item image) & 0.4539 & 0.3370 & 0.2818 & 0.3576 & 0.2465 & 0.2825 & 0.4246 & 0.3179 & 0.3377 \\
RFPR (sort by view count) & 0.6632 & 0.5616 & 0.5226 & 0.5825 & 0.4456 & 0.5002 & 0.7607 & 0.5688 & 0.5757 \\
RFPR (sort by selling count) & 0.6248 & 0.5113 & 0.4672 & 0.5344 & 0.5029 & 0.5453 & \underline{0.7682} & 0.6055 & 0.5700 \\
RFPR (sort by manually-tuned formula) & 0.6552 & 0.5421 & 0.4997 & 0.5657 & 0.4974 & 0.5422 & 0.7626 & 0.6007 & 0.5832 \\
RFPR (sort by manually-tuned + IRF$^\dagger$ \cite{jung2007clickAsImplicitRelevanceFeedback, chakrabarti2008ADwithClickFeedback}) & 0.6579 & 0.5437 & 0.5009 & 0.5675 & 0.4985 & 0.5431 & 0.7629 & 0.6015 & 0.5845 \\
\midrule
HPO (random; 500 samples) & 0.7205 & 0.5817 & 0.5296 & 0.6106 & 0.4958 & 0.5392 & 0.7613 & 0.5988 & 0.6047 \\
HPO (grid- extracted as ctr metrics best) & \underline{0.7326} & \underline{0.5902} & \underline{0.5367} & \underline{0.6198} & 0.4999 & 0.5429 & 0.7649 & 0.6026 & \underline{0.6112} \\
HPO (grid- extracted as ctcvr metrics best) & 0.7188 & 0.5817 & 0.5296 & 0.6100 & \underline{0.5083} & \underline{0.5503} & 0.7645 & \underline{0.6077} & 0.6089 \\
\midrule
$\textit{(Single Objective Tune)}$ & \\
MOHPER (TPE- ctr) & 0.7312 & 0.5851 & 0.5300 & 0.6154 & 0.5029 & 0.5454 & 0.7598 & 0.6027 & 0.6091 \\
MOHPER (TPE- ctcvr) & 0.7293 & 0.5815 & 0.5261 & 0.6123 & 0.5166 & 0.5567 & 0.7646 & 0.6127 & 0.6125 \\
$\textit{(Separate Multi-Objective Tune)}$ & \\
MOHPER (TPE- ctr+ctcvr) & 0.7328 & 0.5875 & 0.5327 & 0.6177 & 0.5169 & 0.5575 & 0.7674 & 0.6139 & 0.6158 \\
MOHPER (TPE- ctr+ctcvr+ctcar) & 0.7315 & 0.5881 & 0.5342 & 0.6179 & \textbf{0.5184} & \textbf{0.5584} & 0.7668 & \textbf{0.6145} & 0.6162 \\
MOHPER (TPE- ctr+ctcvr+ctcar; 500 samples) & \textbf{0.7431} & \textbf{0.5950} & \textbf{0.5395} & \textbf{0.6259} & 0.5167 & 0.5570 & 0.7672 & 0.6136 & \textbf{0.6197} \\
\textit{(Weighted Sum-Based Multi-Objective Tune)} & \\
MOHPER (TPE- ctr+ctcvr) & \textbf{0.7400} & \textbf{0.5941} & \textbf{0.5392} & \textbf{0.6244} & 0.5130 & 0.5539 & 0.7674 & 0.6114 & 0.6179 \\
MOHPER (TPE- ctr+ctcvr+ctcar) & 0.7228 & 0.5790 & 0.5251 & 0.6090 & \textbf{0.5232} & 0.5613 & 0.7678 & 0.6175 & 0.6132 \\
MOHPER (TPE- ctr+ctcvr+ctcar; 500 samples) & 0.7388 & 0.5926 & 0.5379 & 0.6231 & 0.5229 & \textbf{0.5618} & \textbf{0.7683} & \textbf{0.6176} & \textbf{0.6204} \\

\bottomrule
\end{tabular}
}

\centering
\begin{tablenotes}
    \small
    \centering
    \item $^{\dagger}$: Humans manually decide the weights of each model, and choose CTR-based implicit relevance feedback (IRF) to be added for ranking.
\end{tablenotes}

\end{table*}
\begin{table*}[h!]
\caption{Performance comparison of various models and optimization objectives on CPLP test data.}
\label{CPLP_offline_result}
\centering
\resizebox{.85\textwidth}{!}{%
\begin{tabular}{l||cccc:cccc|c}
\toprule
\textbf{Method} & \makecell{\textbf{CTR} \\ \textbf{nDCG@20}} & \makecell{\textbf{CTR} \\ \textbf{nDCG@100}} & \makecell{\textbf{CTR} \\ \textbf{prec@100}} & \makecell{\textbf{CTR} \\ \textbf{avg metrics}} & \makecell{\textbf{CTCVR} \\ \textbf{nDCG@20}} & \makecell{\textbf{CTCVR} \\ \textbf{nDCG@100}} & \makecell{\textbf{CTCVR} \\ \textbf{prec@100}} & \makecell{\textbf{CTCVR} \\ \textbf{avg metrics}} & \makecell{\textbf{Overall} \\ \textbf{avg metrics}} \\
\midrule
RFPR (sort by view count) & 0.6323 & 0.6673 & \underline{0.7966} & 0.6987 & 0.5014 & 0.5400 & 0.8613 & 0.6342 & 0.6665 \\
RFPR (sort by sell count) & 0.5665 & 0.5471 & 0.5803 & 0.5646 & \underline{0.5893} & \underline{0.6201} & \underline{0.8697} & \underline{0.6930} & 0.6288 \\
RFPR (sort by manually-tuned formula) & 0.6263 & 0.6551 & 0.7786 & 0.6867 & 0.5514 & 0.5870 & 0.8672 & 0.6685 & 0.6776 \\
\midrule
HPO (random; 500 samples) & 0.6340 & 0.6605 & 0.7806 & 0.6917 & 0.5680 & 0.6016 & 0.8692 & 0.6796 & \underline{0.6857} \\
HPO (grid- extracted as ctr metrics best) & \underline{0.6436} & \underline{0.6697} & 0.7885 & \underline{0.7006} & 0.5480 & 0.5831 & 0.8668 & 0.6660 & 0.6833 \\
HPO (grid- extracted as ctcvr metrics best) & 0.6305 & 0.6559 & 0.7767 & 0.6877 & 0.5679 & 0.6010 & 0.8688 & 0.6793 & 0.6835 \\
\midrule
$\textit{(Single Objective Tune)}$ & \\
MOHPER (TPE- ctr) & \textbf{0.6514} & \textbf{0.6764} & \textbf{0.7912} & \textbf{0.7063} & 0.5502 & 0.5835 & 0.8655 & 0.6664 & 0.6864 \\
MOHPER (TPE- ctcvr) & 0.6235 & 0.6522 & 0.7740 & 0.6832 & 0.5744 & 0.6058 & \textbf{0.8723} & \textbf{0.6842} & 0.6837 \\
$\textit{(Separate Multi-Objective Tune)}$ & \\
MOHPER (TPE- ctr+ ctcvr) & 0.6321 & 0.6600 & 0.7800 & 0.6907 & 0.5738 & 0.6053 & 0.8722 & 0.6838 & 0.6872 \\
MOHPER (TPE- ctr+ ctcvr+ ctcar) & 0.6374 & 0.6634 & 0.7822 & 0.6943 & 0.5724 & 0.6056 & 0.8693 & 0.6824 & \textbf{0.6884} \\

$\textit{(Weighted Sum Based Multi-Objective Tune)}$ & \\
MOHPER (TPE- ctr+ ctcvr) & 0.6320 & 0.6601 & 0.7800 & 0.6907 & 0.5760 & 0.6072 & \textbf{0.8724} & \textbf{0.6852} & 0.6880 \\
MOHPER (TPE- ctr+ ctcvr+ ctcar) & 0.6377 & 0.6656 & 0.7850 & 0.6961 & 0.5724 & 0.6028 & 0.8718 & 0.6823 & 0.6892 \\
MOHPER (GP- ctr+ ctcvr+ ctcar) & 0.6488 & 0.6712 & 0.7889 & 0.7013 & \textbf{0.5766} & \textbf{0.6073} & 0.8690 & 0.6843 & \textbf{0.6928} \\
\bottomrule
\end{tabular}
}
\end{table*}

\subsubsection{Common Baselines}

We compare ours against the following:

\textbf{BM25}: Lexical sparse retrieval baseline ~\cite{robertson2009bm25}.

\textbf{Dense Retrieval (KoBERT, CLIP)}: Vector-based semantic retrieval from queries to product description (texts or images). For Korean, we employ KoBERT\footnote{https://github.com/SKTBrain/KoBERT}, a fine-tuned variant of BERT \cite{devlin2018bert} and Jina-clip-v2 \cite{koukounas2024jina-clip-v2}, motivated from \cite{radford2021clip}, with 768-dim.

\textbf{Hybrid Search}: This approach integrates BM25's sparse matching with dense retrieval to enhance ranking performance by leveraging their complementary strengths as \cite{wang2021bertwithBM25, ghawi2019gridhpoknnbm25}.

\textbf{Relevance Filtering with Popularity-based Ranking (RFPR)}: This baseline employs a two-stage ranking strategy. First, items are filtered based on query relevance. Then, the filtered set is ranked according to popularity, measured via historical user interaction metrics such as view/wish/cart/purchase frequencies. RFPR is widely used in e-commerce ranking systems to prioritize high-engagement products, thereby improving overall relevance and user satisfaction. We also evaluate hybrid methods that combine baselines with RFPR.

\subsubsection{Samplers}
We evaluated various HPO strategies as follows:

\begin{itemize}[leftmargin=3pt]

\item \textbf{Random Sampler}: This sampler employs a uniform random search strategy to explore the HP space without any prior knowledge, providing a baseline for comparison.

\item \textbf{Grid Sampler}: The grid search method systematically explores a predefined grid of HP values, ensuring exhaustive evaluation across a fixed set of possible configurations. 

\item \textbf{Gaussian Process (GP) Sampler}: The GP sampler \cite{movckus1975GPSampler} employs Gaussian processes \cite{rasmussen2003GP} to model the objective function and iteratively refine the search space by selecting HP configurations with high expected improvement.

\item \textbf{NSGA-III Sampler}: The NSGA-III sampler \cite{deb2013NSGAIIISampler} is an evolutionary algorithm designed for MOHPO. It applies a non-dominated sorting approach and maintains a diverse set of solutions to balance multiple objectives efficiently.

\item \textbf{Tree-structured Parzen Estimator (TPE) Sampler}: The TPE sampler \cite{bergstra2011TPESampler} uses a probabilistic model to guide the search, building a density estimate of the objective function and using it to propose new configurations.
\end{itemize}

\begin{table*}[ht!]
\centering
\small
\caption{Computational complexity per optimization round for each strategy.}
\centering
\resizebox{.7\textwidth}{!}{%
    \begin{tabular}{|l|l|l|}
    \hline
    \textbf{Method} & \textbf{Complexity} & \textbf{Remarks} \\
    \hline
    TPE (Weighted-sum Based)
    & $\mathcal{O}(k d n + k Q L + n \log n)$ 
    & Sampling: $k d n$; Retrieval + Eval: $k Q L$; Quantile: $n \log n$ \\
    TPE (Separate Multi-objective)
    & $\mathcal{O}(k d n m + k Q (L + m) + n m \log n)$ 
    & Sampling: $k d n m$; Retrieval: $k Q L$; Eval: $k Q m$; Quantile: $n m \log n$ \\
    \hline
    NSGA-III (Weighted-sum Based)
    & $\mathcal{O}(g N d + g N Q L)$ 
    & Sampling: $g N d$; Retrieval + Eval: $g N Q L$ \\
    NSGA-III (Separate Multi-objective)
    & $\mathcal{O}(g N d + g N Q (L + m) + g N^2 m)$ 
    & Sampling: $g N d$; Retrieval: $g N Q L$; Eval: $g N Q m$; Sorting: $g N^2 m$ \\
    \hline
    GP (Weighted-sum Based)
    & $\mathcal{O}(k d n^2 + k Q L + n^3)$ 
    & Sampling: $k d n ^2$; Retrieval + Eval: $k Q L$; GP train: $n^3$ \\
    \hline
    \end{tabular}%
}

\begin{tablenotes}
    \small
    \centering
    \item 
    - $n$: \# of evaluated trials (observations), 
    $d$: dimension of the HP space,
    $m$: \# of objectives,
    $k$: \# of sampled candidates per iteration
    \item - For NSGA-III, 
    $g$: \# of generations, 
    $N$: Population size per generation, 
    $Q$: \# of queries used in train/eval,
    $L$: Abbreviations of costs in search engines
\end{tablenotes}

\label{computational_cost}
\end{table*}

\subsection{Experimental Results and Discussion}

\subsubsection{RQ1a. Offline Performance Comparison on Balancing Multi-objective E-commerce Metrics} \label{RQ1a}

To assess the efficacy of the proposed MOHPER framework, we benchmark its performance against several baseline approaches. Specifically, we examine its application in two distinct e-commerce scenarios: retrieval with keyword queries for search engine result pages (SERP) and retrieval based on category IDs for product list pages (CPLP).
The experimental setup involves optimizing three primary metrics (nDCG@20, precision@100) and varying relevance targets across the following five multi-objective scenarios:
(a) CTR (2 objectives), (b) CTCVR (2 objectives), (c) CTR+CTCVR (4 objectives), (d) CTR+CTCVR+CTCAR (6 objectives).
For multi-objective training, we adopt two approaches: (i) separate multi-objective tuning, where each objective is trained individually within a multi-objective framework, and (ii) weighted sum-based multi-objective tuning, which aggregates metrics into a single weighted objective for optimization. We limit HPO methods to 100 samples by default to ensure fairness. For a broader comparison, we also evaluate further sampled configurations. Grid search baselines explore 1,024 configurations for SERP and 512 for CPLP.

The comparative results, presented in Table \ref{SERP_offline_result} and Table \ref{CPLP_offline_result}, demonstrate that MOHPER consistently outperforms baseline methods such as RFPR, random search, and grid search across all configurations. While single-objective approaches (a and b) achieve strong performance for their respective primary objectives (e.g., CTR in scenario a), they fail to optimize complementary metrics effectively (e.g., CTCVR in scenario a).
In contrast, multi-objective configurations, particularly scenario (c), exhibit a more balanced optimization across CTR and CTCVR, yielding superior overall performance. Furthermore, the inclusion of auxiliary objectives in scenarios (d) and (e) enhances the robustness of the results, enabling more generalized performance across diverse evaluation metrics. These findings highlight the effectiveness of MOHPER in addressing complex multi-objective optimization problems in e-commerce retrieval systems.

\subsubsection{RQ1b. Performance Difference Based on Samplers and Setups}

\begin{figure}[!thb]
  \centering
  \includegraphics[width=.85\linewidth]{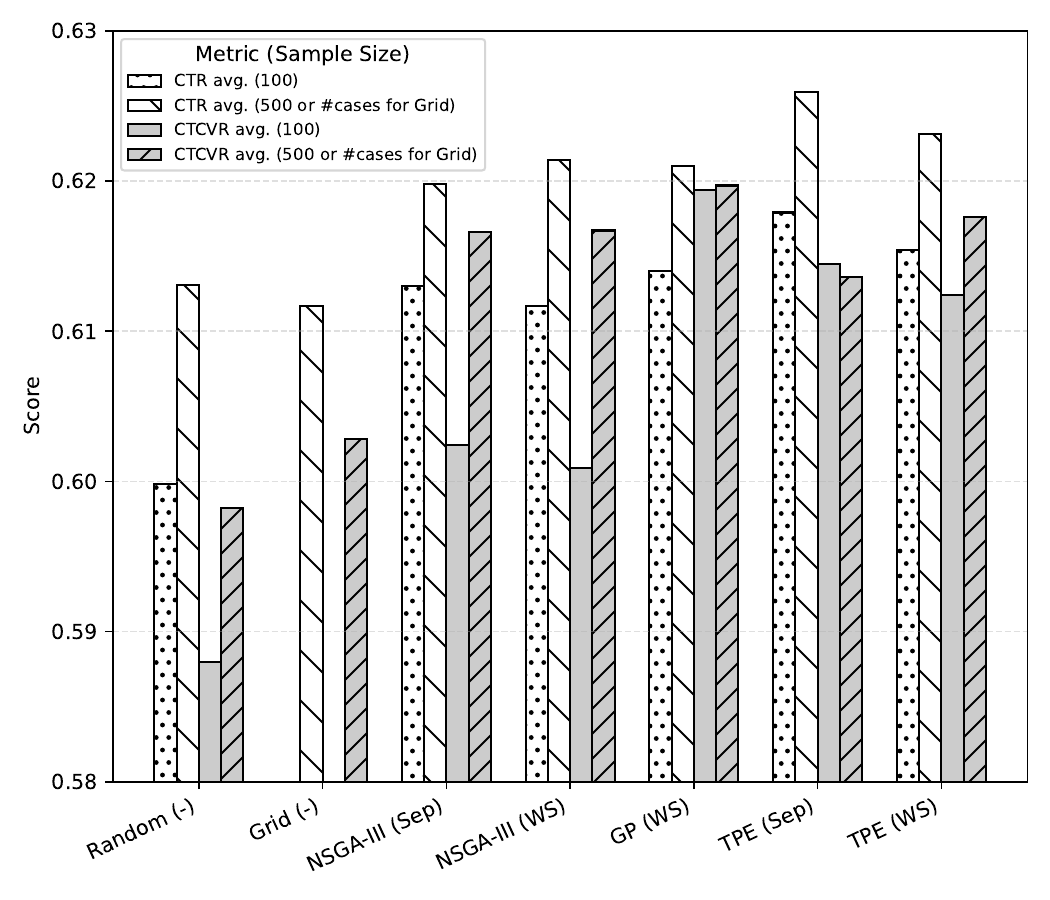}
  \caption{Performance differences across various sampler configurations on the SERP dataset.}
  \label{sampler_results}
  \Description{result_serp_sampler_compare}
\end{figure}

To evaluate the effectiveness of different samplers, we compared their performance across varying sample sizes to determine their suitability. The experimental setup involved training the same objectives outlined in \ref{RQ1a}, while varying the type of samplers and the number of samples. The comparative results are summarized in Figure \ref{sampler_results}., and the computational costs are described in Table \ref{computational_cost}.

While sampler complexity varies, retrieval and evaluation dominate the computational cost. Thus, sampler choice has a trivial impact on performance, while \# of samples, search space, and objective definitions play a more significant role. Among the evaluated samplers, the TPE sampler consistently demonstrated superior performance across all cases and proved to be applicable in diverse scenarios. In contrast, the GP sampler excelled in settings with a small number of samples; however, it is inherently limited to single-objective optimization unless multi-objective problems are reformulated into a single-objective using weighted summation.
The NSGA-III sampler, while applicable to multi-objective scenarios without such reformulation, exhibited reduced effectiveness when the number of samples was constrained.
The results indicate that increasing the number of samples generally enhances the performance of samplers, albeit at the cost of higher computational overhead.
To achieve a more efficient trade-off between performance and computational cost, we introduced a cumulative training strategy, as detailed in the experiments of \ref{rq3}. This approach leverages incremental sampling and training to boost performance while maintaining computational efficiency.

\subsubsection{RQ2. Hybrid Retrieval Arbitration} \label{rq3}

Modern e-commerce systems employ complex retrieval strategies that involve a wide range of feature selection choices, encompassing both sparse (lexical) and dense (vector-based) embedding retrieval methods and item popularity. To illustrate the interplay among three core feature types in e-commerce retrieval, sparse lexical similarity, dense semantic similarity, and item popularity, we utilize BM25, KoBERT, and business quantity features of items, respectively.
BM25 represents sparse lexical similarity, while LLM captures dense semantic similarity. Additionally, business quantity features highlight item popularity, a critical aspect in real-world e-commerce retrieval. To underscore the significance of popularity-based features over similarity-based features in production systems, we also include RFPR baselines.

The evaluation metrics, nDCG@20 and Recall@100, are calculated on query-item pairs using CTR or CTCVR as relevance labels. The results, summarized in Table \ref{retrieval_strategies}, demonstrate that the proposed MOHPER framework effectively balances the contributions of sparse and dense similarities with popularity features. This adaptability enables MOHPER to align retrieval strategies with user preferences and business objectives, optimizing performance across diverse e-commerce scenarios.

\begin{table}[h!]
\caption{Performance change as retrieval strategies on SERP}
\label{retrieval_strategies}
\small
\centering
\resizebox{.9\columnwidth}{!}{%
    \centering
    \begin{tabular}{l|cccc:c}
    \toprule
    \textbf{Method} & 
    \makecell{\textbf{CTR} \\ \textbf{nDCG@20}} & 
    \makecell{\textbf{CTCVR} \\ \textbf{nDCG@20}} & 
    \makecell{\textbf{CTR} \\ \textbf{recall@100}} & 
    \makecell{\textbf{CTCVR} \\ \textbf{recall@100}} & 
    \makecell{\textbf{Overall} \\ \textbf{Avg.}} \\
    \midrule
    BM25 & 0.2270 & 0.0831 & 0.2516 & 0.3951 & 0.2392 \\
    \makecell[l]{ANN \\- (w/ KoBERT)} & 0.3753 & 0.3227 & 0.2270 & 0.4604 & 0.3463 \\
    \makecell[l]{ANN + BM25 \\ - (grid search)} & 0.3855 & 0.3502 & 0.2139 & 0.4551 & 0.3512 \\
    \makecell[l]{RFPR \\- (by selling)} & 0.6248 & \underline{0.5029} & 0.4108 & \underline{0.7682} & 0.5767 \\
    \makecell[l]{ANN + RFPR} & \underline{0.6253} & 0.5026 & \underline{0.4125} & 0.7681 & \underline{0.5771} \\
    \midrule
    \makecell[l]{MOHPER}  & \textbf{0.7400} & \textbf{0.5178} & \textbf{0.4718} & \textbf{0.7685} & \textbf{0.6245}  \\
    \bottomrule
    \end{tabular}%
}
\end{table}

\subsubsection{RQ3. Effectiveness of Voting and Cumulative Learning} \label{rq3}

\begin{table}[!]
\centering
\caption{Performance change across the cumulative training stage with ablation analysis of voting in the meta-stage}
\label{result_cumulative_learning}
\small
\resizebox{.9\columnwidth}{!}{%
\centering
\begin{tabular}{c|cc:cc}
\toprule

\multicolumn{1}{c}{} & \multicolumn{2}{c}{\textbf{Separate Objectives Tune}} & \multicolumn{2}{c}{\textbf{Weighted Sum-based}} \\ 
\cmidrule(lr){2-3} \cmidrule(lr){4-5}

\makecell[c]{\textbf{Cumulative Stage} \\ \textbf{(\# sample/\# total)} \\ \textbf{(default: w/ voting)} }  & \makecell{\textbf{on SERP} \\ \textbf{avg.} \\ \textbf{metrics}} & \makecell{\textbf{on CPLP} \\ \textbf{avg.} \\ \textbf{metrics}} & \makecell{\textbf{on SERP} \\ \textbf{weighted} \\ \textbf{score}} & 
\makecell{\textbf{on CPLP} \\ \textbf{weighted} \\ \textbf{score}} \\ 
\midrule
\makecell[c]{[Stage 0; 500/500]\\ (wo/ voting)} &  0.6221 & 0.6759 & 0.9365 & 1.0884 \\
\makecell[c]{[Stage 0; 500/500]}           & \underline{0.6257} & \underline{0.6778} & \underline{0.9374} & \underline{1.0884} \\

\midrule

\makecell[c]{[Stage 0; 100/100]}           & 0.6223 & 0.6756 & 0.9344 & 1.0762 \\
\makecell[c]{[Stage 1; 100/200]}           & 0.6249 & 0.6756$^\dagger$ & 0.9344$^\dagger$ & 1.0920 \\
\makecell[c]{[Stage 2; 100/300]}         & \textbf{0.6257} & \textbf{0.6786} & 0.9354 & \textbf{1.0920}$^\dagger$ \\
\makecell[c]{[Stage 3; 100/400] \\ (wo/ voting)}   & 0.6244      & 0.6781 & 0.9354     & 1.0864 \\
\makecell[c]{[Stage 3; 100/400]}            & \textbf{0.6257}$^\dagger$      & \textbf{0.6786}$^\dagger$ & \textbf{0.9370}  & \textbf{1.0920}$^\dagger$ \\
\bottomrule
\end{tabular}
}
\centering

\begin{tablenotes}
    \small
    \item $^\dagger$: No incremental performance is observed from the previous stage, likely due to stochastic variations in sampling logic.
\end{tablenotes}

\end{table}

This section demonstrates the effectiveness of voting and cumulative learning strategies in optimization. The results, presented in Table \ref{result_cumulative_learning}, indicate that voting mechanisms can improve performance compared to naive optimization. This approach, consistent with the benefits observed in ensemble methods \cite{chan1995voting, bauer1999empiricalVoting, dietterich2000ensemble}, reduces variance and enhances generalization.
However, cumulative learning further improves performance by iteratively building on prior knowledge, which reduces the need for large sample sizes and minimizes redundant noise introduced by samplers. We hypothesize that this approach effectively mitigates redundant information, particularly in samplers such as TPE, which does not explicitly account for variance during its optimization process. As training progresses, the average differentiation between high- and low-performing samples diminishes due to convergence.
Cumulative learning counteracts this effect by selectively retaining high-quality samples and filtering out redundant or low-quality ones, thus restoring differentiation and enabling more efficient exploration of the HP space.
Specifically, for TPE, this refined focus on promising regions enhances its ability to identify optimal configurations, leading to uplift performance. Moreover, an additional advantage of splitting \# samples into a series of mini-cumulative stages is improved debuggability, and the simplicity of implementation between stages further enhances the practicality and effectiveness of this approach.

\subsubsection{RQ4. Online AB Test}


We present empirical results from applying our proposed methods to various e-commerce experiments on OHouse retrieval tasks. To isolate the performance of MOHPER, we maintained the original search space from previous setups and focused solely on adjusting service hyperparameters. These adjustments included modulating model/feature weights and refining model/feature selections for retrieval or ranking.
To ensure robust estimation, we employed cumulative training over three stages with 500 sampled configurations per stage.
Each round of HPO was conducted with a single compute node with three parallel worker processes.
The total optimization process with this setting required approximately 18 hours per stage (total 2 days) on the SERP domain dataset (comprising 5,000 training queries) and 7 hours per run on the CPLP domain (total 21 hours). 
Notably, our search space was carefully designed to avoid introducing latency bottlenecks. As such, the deployment of new configurations did not negatively impact the stability or responsiveness of the production system. 

The experiments on SERP were conducted over two weeks, while those on CPLP  spanned three weeks. Table \ref{online_exp}. summarizes the results, focusing on key performance metrics: CTR, CTCAR, CTCVR, and per-query engagement rates (Click/Q, Cart/Q, and Purchase/Q). While CTR on CPLP showed a slight decline, all other metrics exhibited consistent improvements, demonstrating MOHPER’s effectiveness in optimizing retrieval performance across multiple objectives. These gains were achieved without increasing computational costs and introducing additional features.
Although we do not provide additional online empirical results for fairness and clarification, we note that subsequent experiments expanding the search space were successfully launched following these conditioned successes.

\begin{table}[]
\caption{Online empirical results of applying MOHPER to SERP and CPLP on OHouse retrieval tasks, meaning $^* : (p < 0.1), ^{**}: (p < 0.05), $ and $^{***}: (p < 0.01) $}
\label{online_exp}
\centering
\resizebox{.9\columnwidth}{!}{%
    \begin{tabular}{l|cccccc}
    \toprule
    \textbf{Exp.} & \makecell{CTR\\ gain(\%)} &  \makecell{CTCAR\\ gain(\%)} & \makecell{CTCVR\\ gain(\%)} & \makecell{Click/Q\\ gain(\%)} & \makecell{Cart/Q\\ gain(\%)} & \makecell{Purchase/Q\\ gain(\%)} \\
    \midrule
    \makecell{on SERP} & 0.95\%*** & 0.75\%*** & 0.30\% & 1.23\%*** & 0.98\%* & 0.75\%* \\
    \makecell{on CPLP} & -0.51\%*** & 1.16\%*** & 1.10\%* & -0.22\%** &1.45\%** & 1.62\%* \\
    \bottomrule
    \end{tabular}%
}
\end{table}

\section{Conclusions}

This paper presents MOHPER, a novel multi-objective HPO framework tailored to the demands of e-commerce retrieval, specifically optimizing for critical performance metrics such as CTR, CTCAR, and CTCVR, which are pivotal for enhancing user engagement and conversions in e-commerce systems. As the first multi-objective HPO framework applied to retrieval and ranking tasks in this domain, MOHPER effectively tackles the inherent complexity of concurrently optimizing diverse and competing metrics in e-commerce.
Through extensive evaluation, we demonstrated MOHPER’s ability to achieve superior stability, convergence, and robustness in multi-objectives optimization across various HPO sampling strategies, achieving significant improvements in CTR, CTCVR, and related objectives (RQ1).
The integration of hybrid optimization mechanisms further enhanced MOHPER’s performance, delivering measurable gains in relevance and user engagement, thus advancing multi-objective optimization for e-commerce (RQ2).
Additionally, we developed advanced techniques such as meta-configuration voting and cumulative training, which uplift MOHPER’s adaptability by shortening sampling costs (RQ3) and also provide inspection changes between each cumulative stage.
Lastly, large-scale A/B testing conducted on OHouse’s dataset demonstrated MOHPER’s scalability and validated its practical utility, solidifying its role as a foundational optimization tool within the platform (RQ4).




\bibliographystyle{ACM-Reference-Format}
\bibliography{ref}

\newpage

\end{document}